\documentclass{aa}  

\usepackage{graphicx}
\usepackage[normalem]{ulem}
\usepackage{txfonts}
\begin{document} 

    % Title and subtitle of first submission
    %\title{A 4,000 au streamer infalling onto the protoplanetary disk of M512.}
    %\subtitle{Dust properties and mass infall rate with two ALMA bands.}

    % resubmission title/subtitle?
    \title{A dusty streamer infalling onto the disk of a class I protostar.}
    \subtitle{ALMA dual-band constraints on grain properties and mass infall rate.}

    \titlerunning{A 4,000 au dusty streamer in Orion}

   \author{L. Cacciapuoti
          \inst{1,2,3}
          \and
          E. Macias 
          \inst{1}
          \and
          A. Gupta
          \inst{1}
          \and
          L. Testi 
          \inst{4}
          \and 
          A. Miotello
          \inst{1}
          \and
          C. Espaillat
          \inst{5}
          \and 
          M. K\"{u}ffmeier
          \inst{6}
          \and
          S. van Terwisga
          \inst{7}
          \and 
          J. Tobin
          \inst{8}
          \and
          S. Grant
          \inst{6}
          \and
          C. F. Manara
          \inst{1}
          \and
          D. Segura-Cox
          \inst{9}
          \and
          J. Wendeborn
          \inst{5}
          \and
           R. S. Klessen
          \inst{10,11}
          A. J. Maury
          \inst{12}
          \and
          U. Lebreuilly 
          \inst{12}
          \and
          P. Hennebelle 
          \inst{12}
          \and
          S. Molinari
          \inst{13}
          }

   \institute{European Southern Observatory, Karl-Schwarzschild-Strasse 2 D-85748 Garching bei Munchen, Germany
   \and
   Fakultat f\'{u}r Physik, Ludwig-Maximilians-Universit\"{a}t M\"{u}nchen, Scheinerstra{\ss}e 1, 81679 M\"{u}nchen, Germany
   \and
   INAF, Osservatorio Astrofisico di Arcetri, Largo E. Fermi 5, 50125, Firenze, Italy  
   \and
   Dipartimento di Fisica e Astronomia "Augusto Righi" Viale Berti Pichat 6/2, Bologna
   \and
    Institute for Astrophysical Research, Department of Astronomy, Boston University, 725 Commonwealth Avenue, Boston, MA 02215, USA; cce@bu.edu
   \and
    Max-Planck Institut für Extraterrestrische Physik (MPE), Giessenbachstr. 1, D-85748, Garching, Germany
   \and
    Max-Planck-Institut f\"{u}r Astronomie, K\"{o}nigstuhl 17, Heidelberg, Germany
   \and
   National Radio Astronomy Observatory, 520 Edgemont Rd., Charlottesville, VA 22903, USA
   \and
    University of Texas at Austin, Department of Astronomy, 2515 Speedway, Stop C1400, Austin, Texas 78712-1205
    \and
    Universität Heidelberg, Zentrum f\"{u}r Astronomie, Institut f\"{u}r Theoretische Astrophysik, Albert-Ueberle-Straße 2, 69120 Heidelberg, Germany
   \and
    Universit\"{a}t Heidelberg, Interdisziplin\"{a}res Zentrum f\"{u}r Wissenschaftliches Rechnen, INF 205, D-69120 Heidelberg, Germany
   \and
   Universit\'{e} Paris-Saclay, Universit\'{e} Paris Cité, CEA, CNRS, AIM, 91191, Gif-sur-Yvette, France
   \and
   INAF-Istituto di Astrofisica e Planetologia Spaziali, Via del Fosso del Cavaliere 100, I-00133, Rome, Italy
}

   \date{Received xx; accepted yy}

% \abstract{}{}{}{}{} 
% 5 {} token are mandatory
 
  \abstract
   %Context
   {Observations of interstellar material infalling onto star- and planet-forming systems have become increasingly common thanks to recent advancements in radio interferometry. These structures have the potential to significantly alter the dynamics of protoplanetary disks, triggering the formation of substructures, inducing shocks, and modifying their physical and chemical properties. Moreover, the protoplanetary disks are replenished with new material, increasing the overall mass budget for planet formation.}
   %Aims
   {In this study, we combine new ALMA band 3 and archival band 6 observations to characterize the dust content and infall rate of a 4,000 au arc-like structure infalling onto [MGM2012] 512 (hereafter M512), a class I young stellar object located in the Lynds 1641 region of the Orion A molecular cloud.}
   %Methods
   {We detect the extended dust emission from this structure in both ALMA bands. We test whether the streamer's velocity pattern is consistent with infalling trajectories by means of analytical streamline models. We measure for the first time spectral index maps and derive a dust opacity index profile along a streamer. We constrain its grain properties and mass.}
   %Results
   {We find that the arc structure is consistent with infalling motions. We measure a spectral index $\alpha \sim$ 3.2 across the entire structure, and a dust opacity index $\beta \sim$ 1.6. Given grain properties consistent with the measured $\beta$, the structure can host up to 245 M$_{\oplus}$ of dust, exceeding or being comparable to the mass of the inner, unresolved 600 au, which contains the protoplanetary disk of M512. Assuming typical ISM dust-to-gas ratio of 1\%, free-fall timescales (50 kyr) imply total mass infall rates up to 1.5 $\cdot$ 10$^{-6}$ M$_{\odot}$/yr. M512 has been classified as an outbursting source with multi-epoch WISE photometry, representing an interesting case study to explore the possible connection between infalling streamers and accretion outbursts.}
   %Conclusions
   {M512 is one-of-a-kind source, for which dust continuum emission of an arc-like streamer extending out to 4,000 au can be characterized in a dual-band analysis. The dust properties results ISM-like, implying a large dust mass. Such a massive streamer can strongly affect the evolution of the star- and planet-forming inner system.}

   \maketitle
%
%-------------------------------------------------------------------

\section{Introduction}
\label{sec:intro}
Star and planet formation takes place in highly dynamical environments in which their accretion history can be far from being driven by the traditionally assumed symmetric collapse. Young stellar objects experience episodic, anisotropic accretion as they travel through the interstellar medium and gravitationally interact with it \citep[e.g.,][]{Lebreuilly2021, Kuffmeier2023}. Modern radiointerferometers have recently revealed large-scale gaseous structures extending from young stars and their protoplanetary disks, at various evolutionary stages. These structures are more commonly observed in near-infrared (NIR) scattered light \citep{Ginski2021} or submillimeter molecular lines, such as CO \citep{Yen2019}, HCO$^+$ \citep{akiyama2019}, H$_2$CO \citep{Valdivia-Mena2022}, HC$_3$N \citep{Valdivia-Mena2023}, among others. Some of them are associated with infall of material, as a stream from the surrounding environment \citep[e.g.,][]{Tobin2010, Alves2020, Yen2014, Yen2019, Garufi2022, Pineda2022}, via gravitational capture of a nearby cloud fragment during a close encounter \citep[e.g.,][]{Scicluna2014, Dullemond2019, Ginski2021, Gupta2023}, or as the result of stripping of material due to stellar flyby events \citep[e.g.,][]{Cabrit2006, Dai2015, Kurtovic2018, Winter2018, menard2020, Dong2022}. 

These so called "streamers" have earned significant interest due to their potential impact on the inner star- and planet-forming system.
Infalling material might induce perturbations in the disk, triggering instabilities (e.g., \citealt{Bae2015}, \citealt{Hennebelle2017}, \citealt{Kuznetsova2022}). These instabilities could then generate exponentially growing vortices, which act as traps for dust particles, where efficient planetesimal formation might begin \citep{Barge1995}. Massive infall can also induce inner-to-outer disk misalignments \citep{Kuffmeier2018, Kuffmeier2021, Ginski2021}, generate spiral waves \citep{Thies2011, Hennebelle2017}, induce accretion outbursts \citep[e.g.,][]{Bonnell1992, Aspin2003} and can result in shocks with the disk material, leading to localized and significant alterations in the physical and chemical conditions of the disk \citep{Garufi2022, Kuznetsova2022}. These variations, in turn, can have a substantial impact on the structural and chemical evolution of the disk. Moreover, the infall of interstellar material can deliver significant amounts of mass to the central protostar and its protoplanetary disk, possibly helping in reconciling the discrepancy between protoplanetary disk masses and exoplanet masses (\citealt{Manara2018}, \citealt{Mulders2021}). The simulations of \citet{Kuffmeier2023} provide an example of how protostars can sweep up material while moving through their natal clouds, and accrete substantial fractions of their final mass in regions even tens of thousands au away from the positions at which we observe them today.
%If such infall events are common during the early phases, they could potentially bridge the observed mass gap between class 0/I and class II disks \citep{Manara2018, Williams2019, Tychoniec2020, Mulders2021}. 
% The observed extended structures can also be caused by stellar flybys, in which a passing object strips material away from the disk of a nearby source \citep[e.g.,][]{Clarke1993, Cuello2019}. Flybys can drive spiral arms in the inner disk of the perturbed objects \citep[e.g.,][]{Clarke1993, Pfalzner2013}, truncate disks \citep[e.g.,][]{manara2019, Zurlo2020}, and induce outbursts \citep[e.g.,][]{Bonnell1992, Aspin2003}. A recent, comprehensive review of flyby physics and its implications can be found in \citet{Cuello2023}. Even in this case, at least part of the stripped material is expected to fall back onto the circumstellar disk. 
In summary, infalling streamers are expected to dramatically influence the evolution of the star- and planet-forming systems involved. However, detecting and characterzing examples of such events is critical to constrain their frequency and magnitude, about which little is known so far.
\begin{figure*}[t]  
    \centering
    \includegraphics[width=\linewidth]{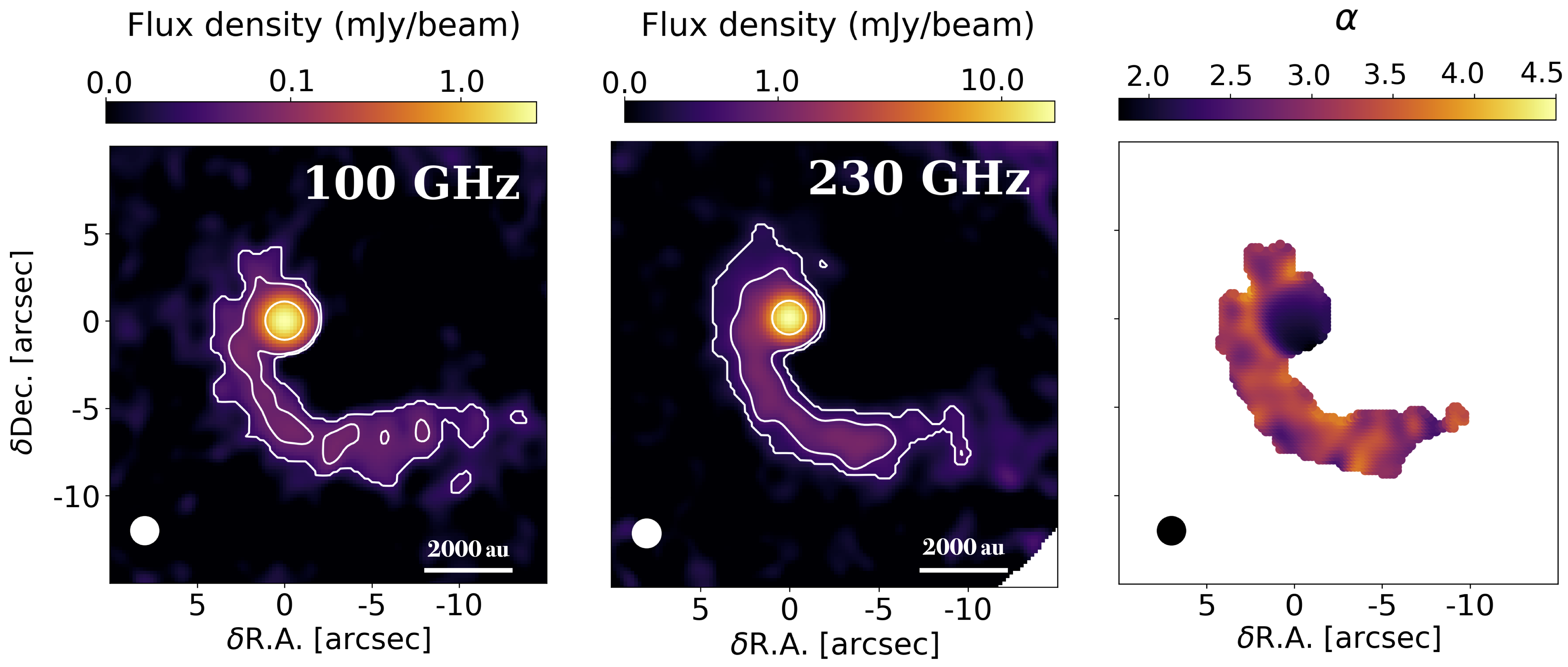}
    \caption{ALMA Band 3 (100 GHz) continuum emission of M512 with [3, 7, 100]$\sigma$ 
    contours overplotted in white (left). ALMA Band 6 (230 GHz) with [3, 10, 100]$\sigma$ contours overplotted in white (center). The spectral index as measured from Eq. \ref{eq:alpha} using only pixels with flux above the 3$\sigma$ noise level (right).
    The x and y axes report offset from the central source in arcseconds. Color bars represent the flux density (mJy/beam) for the ALMA observations (left, center) and the value of the spectral index (right). The images have been created applying a uv-taper to the visibilities, with "briggs" weighting and robust = 1. Then we have smoothed them to the same resolution of 1$\farcs$6, roughly the width of the extended emission. The beam is reported in the lower left.}
    \label{fig:m512}
\end{figure*}

Recently, \citet{grant2021} reported the detection of extended continuum and molecular millimeter emission around [MGM12] 512 (RA $=$ 05h40m13.789s, Dec $=$ -07d32m16.02s). Hereafter, we will refer to the source as M512, consistently with \citet{grant2021}. 
This young source is located in the Lynds 1641 region of the Orion A molecular cloud at a distance of $\sim$ 420 pc, based on the Gaia DR3 parallax p $\sim$ 2\farcs379 \citep{Gaia2022}. This distance is consistent with very long baseline interferometry estimates for Orion Nebula Cluster (e.g., 414$\pm$7 pc, \citealt{Menten2007}). In this work, we assume d = 420 pc.
The source has been classified as class I based on the 2$-$25$\mu$m slope of the spectral energy distribution, $\alpha_{2-25\mu m}\sim0.33$ (\citealt{Caratti2012} and \citealt{Megeath2012}). Finally, its main physical properties have been reported by \citet{Caratti2012} based on near-infrared spectroscopic measurements and we summarize them in Table \ref{tab:m512}.
\begin{table}
\caption{Stellar parameters of M512 as reported by \citet{Caratti2012}, based on its spectral energy distribution and SofI near-infrared spectra.}
\label{tab:m512}
\centering
\begin{tabular}{lccc}
\hline \hline
            Alternative ID  &  IRAS 05377-0733  \\
            RA &  05h40m13.789s \\
            Dec & -07d32m16.02s\\
            SpT & M7.5 $\pm$ 1.5\\
            T$_{eff}$ [K] & 2800 $\pm$ 200 \\
            R$_{*}$ [R$_{\odot}$] & 3.71 \\
            M$_{*}$ [M$_{\odot}$] &  0.15\\
            L$_{*}$ [L$_{\odot}$] &  0.9\\
\hline
\end{tabular}
\end{table}

Here, we present new ALMA Band 3 (B3) observations M512 and combine them with archival ALMA Band 6 (B6) data to constrain the dust properties and mass of the extended structure. In Sect. \ref{sec:obs}, we describe the observations; in Sect. \ref{sec:kinematics} we discuss the infalling nature of the extended arc around M512; and in Sect. \ref{sec:spectral_index}, we measure the spectral index and derive a profile of the dust opacity spectral index. We discuss the results and the possible origins of this structure in Sect. \ref{sec:discuss} and summarize our conclusions in Sect. \ref{sec:conc}.

%--------------------------------------------------------------------

\section{Observations}
\label{sec:obs}
M512 was observed with ALMA in B6 on 2019/11/27 (2019.1.01813.S, PI: van Terwisga, S.) and on 2019/12/12 (2019.1.00951.S, PI: Grant, S.). Both programs achieved a resolution of approximately 1$\farcs$14, and a reported maximum recoverable scale of roughly 11$\farcs$7. We refer the reader to \citet{grant2021} and \citet{vanterwisga2022} for further details and report the main characteristics of their observations in Tab. \ref{table:1}.

We here present new ALMA B3 observations of M512 taken on 2023/01/16 (program 2022.1.00126.S, PI: Wendeborn, J.). These were designed to retrieve the 3 mm continuum flux of the extended emission around M512, first reported by \citet{grant2021} at 1.3 mm, with a similar resolution. The bandpass and flux calibrator is J0423-0120, the phase calibrator is J0542-0913 (see Tab. \ref{table:1}).  

The B3 data were initially calibrated by ALMA staff using the Cycle 9 ALMA interferometric pipeline \citep{Hunter2023} within Common Astronomical Software Applications (CASA; version 6.4.1-12). The B6 data were instead calibrated by the ALMA staff using CASA version 5.6.1-8 for both programs.
We carried out further data reduction and calibration steps in both bands using CASA version 6.2.1.7. First, the ALMA cubes for each execution block were inspected and additional flagging was applied when necessary, i.e. to mask out all spectral lines. Among these, we note that the same arc-like structure is clearly detected in DCO+ (216.11 GHz) and N$_2$D+ (231.32 GHz). These lines are not spectrally resolved and thus could not be used in our kinematic analysis (Sect. \ref{sec:kinematics}).

We then channel-averaged the spectral windows of every execution block. First, we imaged the channel-averaged visibilities with the \textit{tclean} CASA task and fitted a Gaussian with \textit{imfit} to identify the emission peak. Next, we used the \textit{fixvis} function to shift the phase center to the position of the peak of M512 at the epoch of every observation. 
Since we are interested in measuring spectral indices, we checked for potential offsets among the visibility amplitudes of different execution blocks and found none above 3\%. 

We then phase-selfcalibrated the data.
We computed the phase corrections on the continuum spectral windows of each execution block using \textit{gaincal} and applied them with \textit{applycal}. This step aimed to correct phase errors between executions and between spectral windows. We repeated this procedure three times, progressively shortening the solution interval. In the first round, we set the solution interval of \textit{gaincal} to \textit{inf}, combined the scans within each execution block and set gaintype $=$ "G", i.e. the gains were determined for each polarization and spectral window. In the second and third rounds, we combined the spectral windows of each block, shortened the solution intervals to 60 and 10 seconds, respectively, and set gaintype $=$ "T", to obtain one solution for both polarizations. At each round, we split and apply the table computed at the previous steps to each execution block.
The selfcalibration yielded peak S/N improvements of about 15\% as the phase of the visibilities were already well constrained. We did not find any appreciable improvement in the noise and signal-to-noise properties for phase-only selfcalibration steps with even smaller time intervals. The final peak signal-to-noise ratio of the B3 image is 355, and it is 218 for the B6 one.

Finally, we imaged the visibilities to carry out the rest of the analysis.
Since the emission is extended, we ran the \textit{tclean}
routine setting \textit{pbcor = True}, in order to correct for the primary beam attenuation. We deconvolved the images using the "hogbom" algorithm \citep{hogbom74} with the "briggs" weighting scheme, and we tried a number of "robust" parameter choices to find the optimal balance between resolution and sensitivity. We have finally carried out the analysis on images with robust $=$ 1, a pixel size of 0$\farcs$2 (84 au at the distance of M512, one seventh of the synthesized beam) and an image size of 300 pixels. When running \textit{tclean}, we manually masked the emission, comprising of both the bright inner disc and the extended streamer, and CLEANed down to 1.5 times the image rms.
Since we want to compare emission tracing the same physical scales, we used a \textit{uvtaper} to weight the visibilities in order to obtain similar synthesized beams for the two bands (1\farcs5 x 1\farcs2, PA = 72$^{
\circ}$).
Finally, we smoothed the images to the same resolution of 1$\farcs$6, roughly equal to the width of the extended emission at 3$\sigma$ level recovered by \citet{grant2021} with robust = 0.5 (see their Fig. 13).
We recover an arc-like structure radially extending out to about 4,000 au. We note that that the length of the streamer, as observed in this study, is to be considered a lower limit since the interferometric observations might have filtered out larger-scale emission. 
Moreover, we report a C$^{18}$O $J = 2-1$ transition moment-one map of M512 obtained with the observations of ALMA program 2019.1.00951.S (Fig.~\ref{fig:moment1}). 
In what follows, we focus on characterizing the kinematics and the dust content of this extended structure, to which we will refer as "streamer" hereafter.

\section{Streamer's kinematics}
\label{sec:kinematics}
Whether extended structures such as the one linked to M512 are actually infalling on the central protoplanetary disks is a problem that requires precise constraints on the system's geometry and physical properties.
In our case, the unknown details of the geometry of the inner disk, due to somewhat limited spatial resolution, introduce uncertainties in what regards the interpretation of the velocity structure around the source. To complicate the problem further, both the mass and system velocity are uncertain. The former is affected by the precision to which spectral types are determined by near infrared spectra \citep{Caratti2012}, the high extinction of this region (e.g., \citealt{Gutermuth2011}) and the contribution of the inner embedding material connected to an envelope remnant or to the streamer itself.
Furthermore, the limited spectral resolution and the potentially high optical depth of the CO isotopologues limits the accuracy of estimation of the system velocity.
\begin{figure*}[t]
    \centering
    \includegraphics[width=\linewidth]{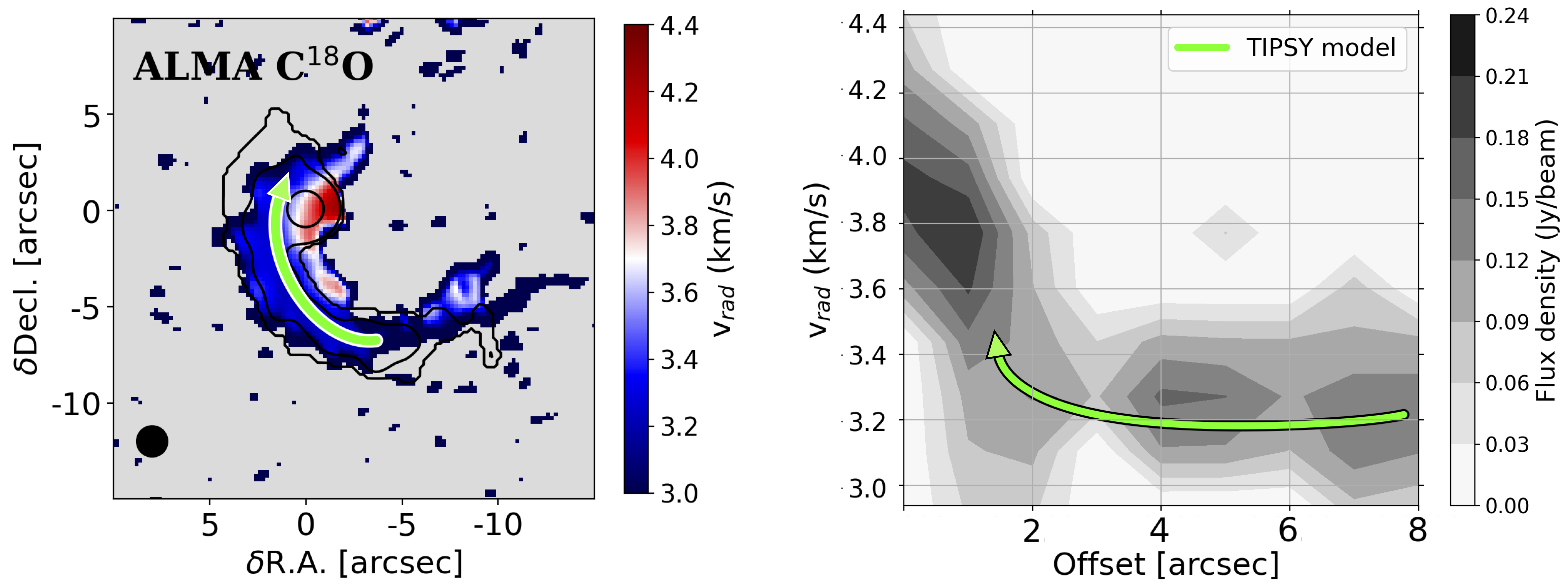}
    \caption{Moment one map of $^{18}$CO J $= 2-1$ transition ($>2 \sigma$) around M512 (\textit{left}). The black contours are the B6 continuum [3, 10, 100]$\sigma$ levels. The green line is the TIPSY model trajectory for M = 0.15 M$_{\odot}$ and v$_{sys}$ = 3.25 km/s. Position-velocity diagram of the streamer's emission (grey contours) extracted along the TIPSY trajectory, and including the inner protostellar region (\textit{right}). The TIPSY fit to the radial velocities is reported as a green line.}
    \label{fig:moment1}
\end{figure*}
In this section, we analyse the C$^{18}$O J=2-1 transition already presented in \citet{grant2021}. Among the other isotopologues, we chose this transition as its emission is optically thinner than the one of $^{12}$CO and $^{13}$CO and less affected by foreground absorption. We imaged the spectral cube containing the line's spectral window with the same scheme of the continuum dataset. During the cleaning, we masked the emission manually, channel-by-channel, in order to capture its complex morphology. 

The moment-one map shows an arc-like feature consistent with the 1.3mm dust continuum emission region, and it is characterized by a progression from lower to higher velocities when approaching the protostar from afar (Fig. \ref{fig:moment1}). This gradient may be indicative of infalling motions towards the inner object (as observed in e.g., Per-emb 2 \citealt{Pineda2020}, Per-emb 50 \citealt{Valdivia-Mena2022}). The velocity transition between streamer and inner material might also indicate rotation of the innermost embedding material.
We attempted to model the C$^{18}$O kinematics to constrain whether the data is consistent with infalling motion. We analysed the spectral cube with the \textit{Trajectory of Infalling Particles in Streamers around Young stars} (TIPSY, Gupta et al., subm.). The input for the code are the mass of the central object and its line-of-sight velocity (v$_{sys}$).
For the former, we set to 0.15 M$_{\odot}$, as reported in Table \ref{tab:m512}. We note that the inner envelope mass, and even the streamer's mass, might contribute significantly to influence the ballistic motion of a particle. 
For the latter, we explored a range of systemic velocities consistent with the observed C$^{18}$O spectral line, between 2.9 and 3.9 km/s.
Based on these constraints, TIPSY solves analytical kinematics equations for an infalling parcel of gas that feels the gravity of a central source from \citet{Mendoza2009} and computes its trajectory along the region over which we detect emission, as well as the kinetic, gravitational (and thus total E$_{tot}$) energy of the parcel and its infall time scale. We overplot a TIPSY fit model to the moment-one map in Fig.\ref{fig:moment1} to show how the arc-like feature around M512 is consistent with the infall of material. The available data is consistent with infall for velocities in the explored range of v$_{sys}$, albeit the material results only loosely bound (E$_{tot} \lesssim$ 0) for v$_{sys} \gtrsim 3.7 km/s$. The TIPSY model constraints the infall time scales to about 50 kyr. We overplot the radial velocities as fitted by TIPSY in a position-velocity diagram (PVD, Fig.\ref{fig:moment1}) obtained using \textsf{pvextractor} \citep{Ginsburg2016}. The PVD is extracted along the infalling trajectory shown in Fig.\ref{fig:moment1}, and including the inner region. The width of the selected path is equivalent to the beam, 1\farcs6. The TIPSY model is consistent with the radial velocities of the observed emission. New observations will be key to constrain the geometry, system velocity and streamer's mass effects on the infall trajectories.

In Sect. \ref{sec:discuss}, we discuss further qualitative arguments that support the infall scenario against alternatives such as stripping due to a flyby, or the cavity wall of an outflow. We thus tentatively suggest that the arc-like structure detected around M512 is an infalling streamer.

\section{Spectral index}
\label{sec:spectral_index}
The slope of the radio spectrum:
\begin{equation}
\mathrm{\alpha} = \frac{\log_{10}{\mathrm{F}(\nu_2)}-\log_{10}{\mathrm{F}(\nu_1)}}{\log_{10}{\nu_2}-\log_{10}{\nu_1}},
\label{eq:alpha}
\end{equation}
where $\nu_i$ are the observed frequencies, can provide valuable insights on the properties of interstellar dust grains. Given that the dust opacity follows a power law relation $\kappa \propto \nu^{\beta}$ in the millimeter end of the spectrum, in the optically thin regime and if the Rayleigh-Jeans (RJ) approximation holds, $\beta = \alpha-2$ (e.g., \citealt{draine2006}, \citealt{BeckwithSargent1991}, \citealt{MiyakeNakagawa1993}, \citealt{Natta2007}).
The typical dust opacity index observed for the interstellar medium, $\beta\sim1.6$, is usually interpreted in terms of maximum grain sizes of the dust population ranging from 100 \AA\ to 0.3 $\mu$m \citep{Weingartner2001}. On the contrary, a value $\beta<1$, which is often measured in class II objects, indicates the presence of larger grains, with sizes $a \geq 1$ mm in more evolved disks (\citealt{BeckwithSargent1991}, \citealt{Ricci2010}, \citealt{Testi2014}, \citealt{Macias2021}).

In the following analysis, we present maps of the spectral index $\alpha$ and the profile of the dust opacity spectral index $\beta$ across the emission region of M512 and its streamer.
% \begin{figure}[t]
%     \centering
%     \includegraphics[width=\linewidth]{pvdiagram_tipsy_model.png}
%     \caption{Position-velocity diagram of the streamer's emission (grey contours) extracted along the TIPSY trajectory of Fig.\ref{fig:moment1} and including the inner protostellar region, with a width of 1\farcs6. The TIPSY fit to the radial velocities is reported as a green line. The dashed line is the centroid of the C$^{18}$O line as extracted from a region centered on the disk/protostar alone.}
%     \label{fig:pvdiag}
% \end{figure}

\subsection{Spectral index maps}
% \begin{figure}[t]
%     \centering
%     \includegraphics[width=\linewidth]{escape_map_inc60.png}
%     \caption{Fraction of C$^{18}$O bound to the inner protostar (M=0.15$M_{\odot}$, v$_{sys}=$3.6km/s). The black contours are the B6 continuum [3, 10, 100]$\sigma$ levels. In this case, we assume an inclination for the arc of sixty degrees.}
%     \label{fig:escape60}
% \end{figure}
\begin{figure*}[h]
    \centering
    \includegraphics[width=\linewidth]{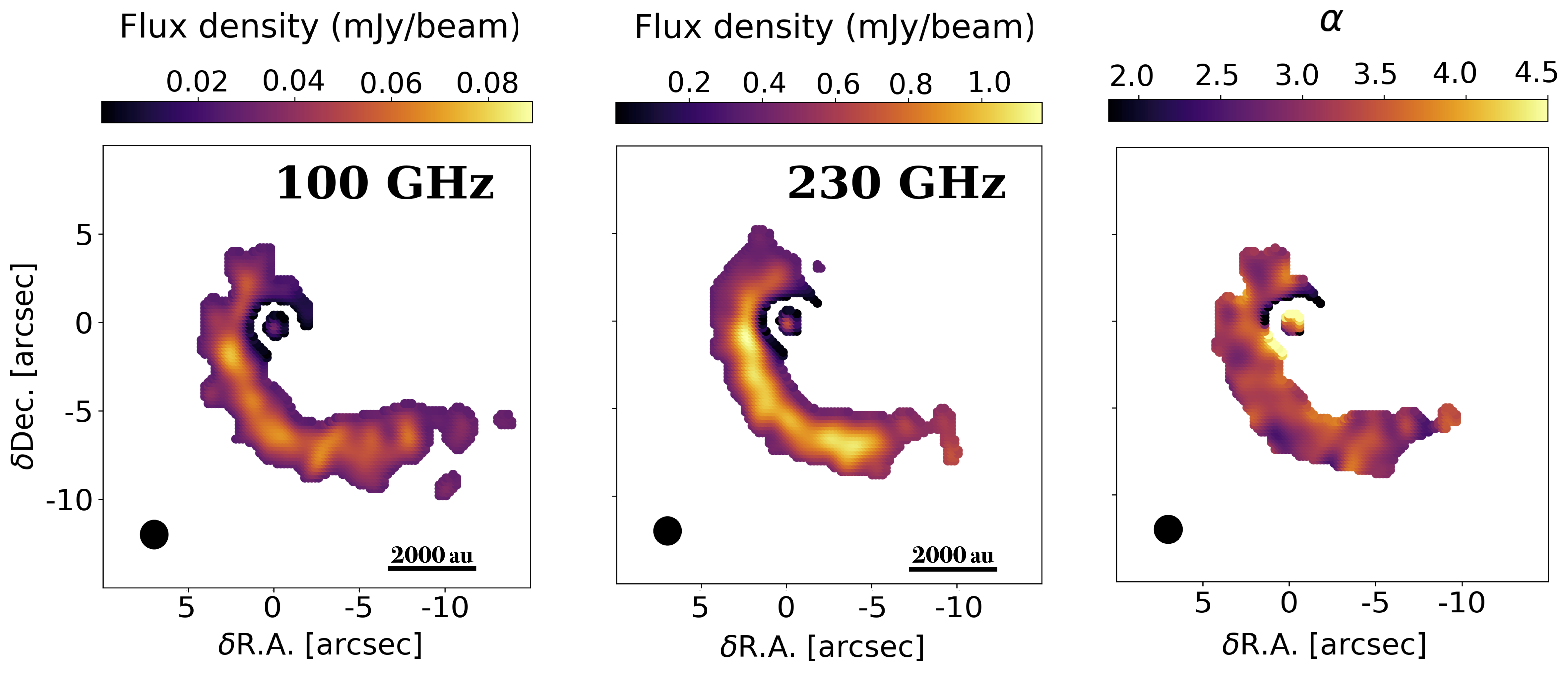}
    \caption{Same as Fig.~\ref{fig:m512}, but after subtraction of the inner, optically thick region. Here, only the pixels with flux above 3$\sigma$ are shown for each inset.}
    \label{fig:streamer}
\end{figure*}
Using the ALMA B3 and B6 maps, we calculate the spectral index $\alpha$ at each pixel in our smoothed maps. This procedure yields the spectral index map shown in Fig.~\ref{fig:m512}. 
Within the inner $\sim$600 au region, centered on the protoplanetary disk of this young source, the spectral index is approximately $\alpha\sim2$, indicating optically thick emission \citep{BeckwithSargent1991}. The high optical depth of this compact region suggests an optically thick disk, diluted on a more extended sky area by the finite resolution of the observations.
To ensure that the spectral index of the inner end of the streamer is not contaminated by this central emission, we model and subtract the contribution of the latter. We fit a 2D Gaussian to the inner region in the image plane. The fitting results are provided in Table \ref{tab:gaussfit}, and the model and residuals are displayed in Appendix \ref{app:gaussfit}. The compact emission extends for 1$\farcs$4 (2$\sigma$, 574 au), possibly comprising a combination of the inner end of the streamer, remnants of the original envelope of M512 and a more compact protoplanetary disk. Higher resolution observations will be key to discern the extent of the inner disk.
We then subtract the Gaussian model from the image to isolate the contribution of the streamer (Fig.~\ref{fig:streamer}). Finally, we compute the spectral index of the streamer emission, $\alpha_{S}$ (refer to Fig.~\ref{fig:m512} and Fig.~\ref{fig:streamer}). We observe a spectral index of approximately $\alpha_{S}\sim3.2$ along the entire streamer, which indicates optically thin emission. Given the dust temperature profile we consider to derive the dust opacity spectral index (see later Eq.\ref{eq:alpha}), we derive $\tau = -\ln(1 - I_{\nu}/B_{\nu}(T)) < 0.05$ in B6 along the full arc.
As a sanity check, we ran a similar analysis in the uv plane, where we subtracted a point source from the visibilities and then computed the spectral index as a function of uv-distance. We obtain consistent results (see Appendix \ref{app:gaussfit}).

% \begin{figure*}
%     \centering
%     \includegraphics[width=\linewidth]{spectral_index_map_streamer.png}
%     \caption{Spectral index map of M512 total emission computed using ALMA Band 3 and Band 6 data (left). The compact region is optically thick ($\alpha \sim 2$). The spectral index map of the streamer, after subtraction of the compact emission is shown on the right and reveals an optically thin structure where $\alpha_{S} \sim 3.2$. The axes and beam are as in Fig.~\ref{fig:m512}.}
%     \label{fig:alpha_maps}
% \end{figure*}

\subsection{Dust opacity index profile}

To investigate if and how dust optical properties change along the extended arc of M512, we divide the streamer in spatial bins and derive the dust opacity spectral index $\beta$ within each bin. In order to select the bins, we use a logarithmic spiral with manually-tuned parameters to adapt it to the shape of the extended emission. We then select points along the spiral as centers for circular apertures as large as the streamer's width (see Fig.~\ref{fig:beta_profile}). The aperture's radius is twice the beam's one and each aperture contains roughly 160 pixels.
If we assume that the main heating mechanism for the dust is the irradiation from the central source, we can use to a first approximation the radial temperature profile that \citet{Motte2001} derived for spherical dusty envelopes:
\begin{equation}
    \mathrm{T(r)} =  38 \mathrm{K} \Bigg( \frac{\mathrm{L}}{\mathrm{L}_{\odot}} \Bigg)^{0.2} \Bigg( \frac{\mathrm{r}}{100 \mathrm{au}} \Bigg)^{-0.4}.
    \label{eq:temp}
\end{equation}
This equation yields temperatures in a range of 21 to 8 K across the scales of the streamer, that is 600 to 4,000 au. 
We note that these distances from the central source assume the streamer lays in the plane of the sky.
At $\lambda =$ 1.3 mm, a temperature T = 8 K implies that the RJ approximation is not valid ($h\nu$/k$_B$T $\sim$ 1). We thus introduce a correction to the simple $\alpha = \beta - 2$ case for the low temperatures which would otherwise artificially lower the derived $\beta$, that we then derive as:
\begin{equation}
   \mathrm{\beta} = \frac{\log \Bigg[ \Big(\mathrm{F}_{\nu_2}(r)/\mathrm{F}_{\nu_1}(r) \Big) / \Big(\mathrm{B}_{\nu_1}(\mathrm{T}_{\mathrm{d}}(r))/\mathrm{B}_{\nu_2}(\mathrm{T}_{\mathrm{d}}(r)) \Big) \Bigg]}{\log(\nu_2 / \nu_1)},
    \label{eq:beta}
\end{equation}
where the ALMA B6 and B3 representative frequecies are $\nu_2 = 230$ GHz, $\nu_1 = 100$ GHz, T$_d$ is the temperature of the dust and B$_{\nu_i}$(T$_d$) is the Planck function value at the two frequencies. 

Finally, we compute the statistical error on the spectral index as:
\begin{equation}
    \delta  \mathrm{\alpha}^2 (= \delta \mathrm{\beta}^2) = \Bigg(\frac{1}{ln{\nu_2}-ln{\nu_1}}\Bigg)^2 \Bigg(\frac{\sigma_1^2}{\mathrm{F}_{\nu_1}^2} + \frac{\sigma_2^2}{\mathrm{F}_{\nu_2}^2}  \Bigg),
\end{equation}
where $\sigma_i$, $F_{\nu_i}^2$  are the primary beam corrected rms and fluxes in each distance bin for the two bands. We consider the same error for $\beta$ since we assume an exact value for the temperature in Eq. \ref{eq:beta}. On top of the statistical error, one should consider ALMA's systematic calibration uncertainty on B3 and B6 fluxes of 1$\sigma$ = 5\%  \citep{Remijan2019}. 
Within the same bins, we also compute $\alpha$ using Eq. \ref{eq:alpha} and show the resulting $\beta = \alpha-2$ proxy and the $\beta$(T) profile in Fig.~\ref{fig:beta_profile}. 

We find a mean $\beta = 1.62 \pm 0.04$ along the whole structure. Finally, we have tested the two extreme scenarios in which the temperature is fixed at 21 or 8 K across the whole structure. In fact, it could be that the arc self-shields against the internal radiation, thus being colder on average or, on the contrary, that significant external irradiation from Orion OB stars could lead to higher temperatures in the outer arc \citep{Haworth2021}\footnote{However, M512 lays in a region where the UV irradiation field is too low to induce such an effect \citep{vanTerwisga2023}.}. 
In any case, the $\beta$ measured with both fixed temperatures only changes by $<$10\% with respect to the one obtained using the temperature profile. 

\section{Discussion}
\label{sec:discuss}
Here, we discuss the interpretation of the derived dust opacity index, we estimate the mass of the surrounding material, and weigh the implications such an event can bear on a planet-forming system, as well as its possible origins.

\subsection{Streamer's dust properties}
\label{sec:discuss_dust}
The retrieved continuum flux of the streamer around M512 implies a relatively large dust mass, that can be distributed either in large amounts of small grains or in smaller amounts of large grains. It is thus important to constrain the maximum grain size of the distribution in order to constrain its mass content.

\begin{figure}[t]
    \centering
    \includegraphics[width=\linewidth]{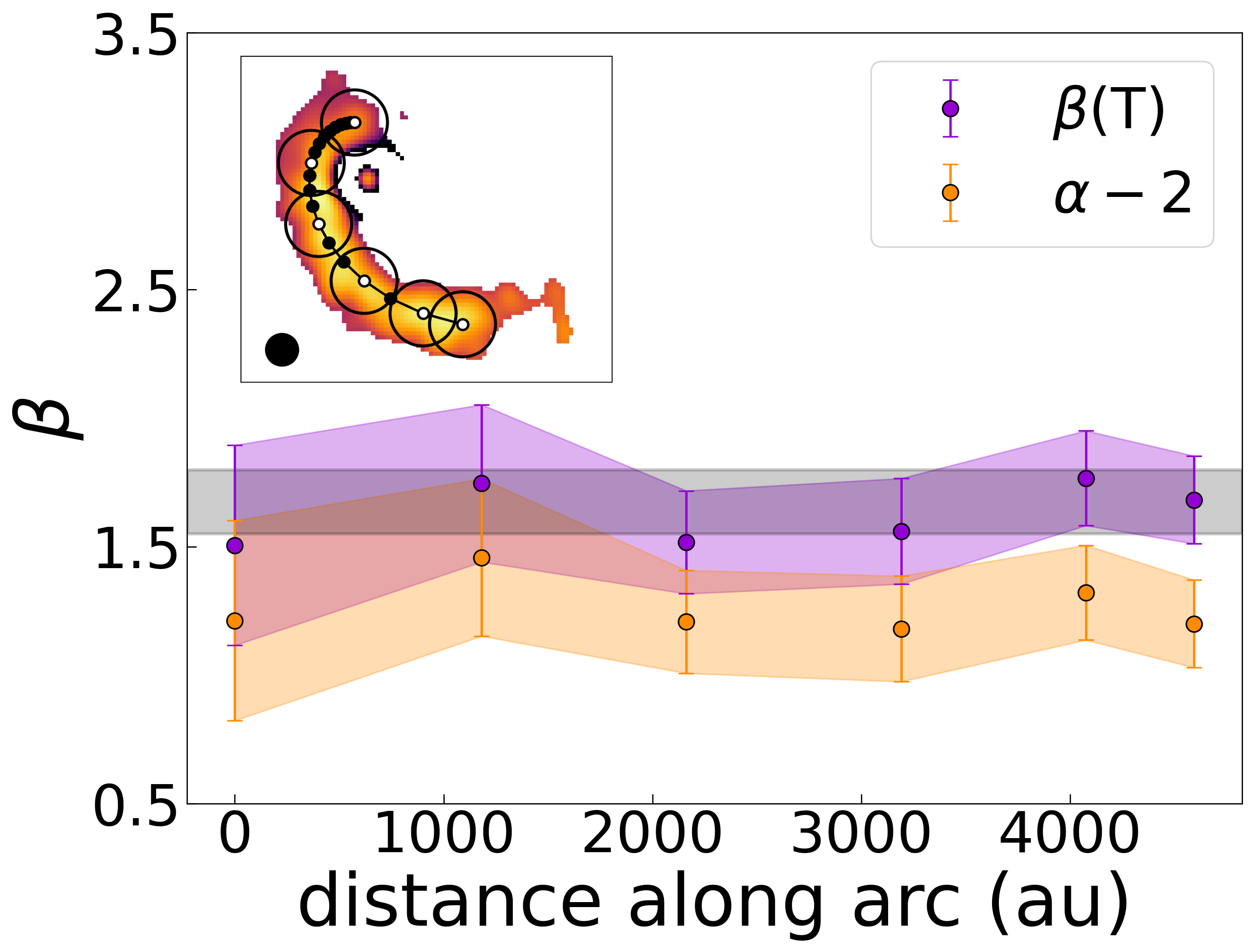}
    \caption{Profile of the dust opacity spectral index $\beta$ along the streamer's length. The approximation $\beta = \alpha-2$ is the orange line while $\beta$, corrected for deviation from RJ approximation, is the violet one. This profile has been obtained starting from the maps in Fig.~\ref{fig:streamer}, where the central emission had been modelled and subtracted. The centers of the selected apertures in which we derive $\beta$ are highlighted in white among all the points that sample the manually-defined logarithmic spiral, on the B6 map as an example (upper inset). Along the whole structure, $\beta\sim1.6$. The grey band represents ISM-like $\beta$.}
    \label{fig:beta_profile}
\end{figure}

In the case of a distribution of spherical compact grains, the Mie scattering theory predicts that the derived $\beta\sim$1.6 (Fig. \ref{fig:beta_profile}) is consistent with a dust population for which the maximum grain size is submicron \citep{draine2006}. If the grains were instead porous, $\beta\sim$1.6 could be consistent with a dust distribution where the maximum grain size lays in a smoother range from submicron to a millimeter (e.g. \citealt{Birnstiel2018}).

Recently, it has been suggested that dust can grow up to millimeter sizes even in protostellar envelopes. In fact, $\beta$ values lower than $\sim$1.6 have been measured in the inner envelopes of a sample of class 0/I objects \citep[e.g.,][]{Kwon09, Miotello2014, Galametz2019, Cacciapuoti2023}. However, models show that dust cannot grow more than a few $\mu$m in more diffuse molecular clouds cores \citep[e.g.,][]{Ormel2009, Lebreuilly2023a}, where the submillimeter dust opacity spectral index $\beta \sim$1.6 \citep{draine2006}. Scattering light observations of these environments have also been independently interpreted as the presence of grains up to only $\sim$ 10 $\mu$m \citep[e.g.,][]{Steinacker2010, Steinacker2014, Steinacker2015}. 
Thus, the $\beta = 1.6$ measured along the streamer of M512 is consistent with the properties of ISM-like dust grains. 

\subsection{Mass constraints and infall rate}
\label{discuss:mass}
The delivery rate of mass onto M512 is critical to assess the implications that the event might have for the subsequent evolution of the system.
In the optically thin regime, the dust mass of the observed streamer can be derived as \citep{Hildebrand1983}:
\begin{equation}
    \mathrm{M} = \frac{\mathrm{d}^2 \mathrm{F}_{\nu}}{\kappa_{\nu} \mathrm{B}_{\nu}\mathrm{(T_d)}}
    \label{eq:mass}
\end{equation}
where d is the distance to the source, F$_{\nu}$ the flux density of the structure, $\kappa_{\nu}$ is the dust opacity and B$_{\nu}$(T$_d$) the value of the Planck function at frequency $\nu$.
We consider a distance of 420 pc (see Section \ref{sec:intro}), and the temperature is given by Eq. \ref{eq:temp}. In order to measure the mass of the streamer taking into account the expected temperature variation along the arc, we consider the logarithmic spiral we mentioned in Sect. \ref{sec:discuss_dust} and shown in Fig.~\ref{fig:beta_profile}. For each selected aperture on the spiral, we assign each pixel above the 3$\sigma$ threshold to the closest aperture center, measure the total flux in the bin so defined, and compute the temperature with Eq. \ref{eq:temp}, where $r$ is the projected radial distance of each aperture center from the source.

%Dust distributions consistent with the $\beta\sim1.6$ we derive would have maximum grain sizes between submicron and a few hundred microns (see previous section). 
Assuming DSHARP dust properties \citep{Birnstiel2018}, the absorption opacity of a dust grains distribution characterized by ISM-like spectral indices as the one we measure for the streamer of M512, and a typical power-law $n(a) \propto a^{-3.5}$, 
%are $\kappa^{a_{max} \gtrsim 100 \mu m}_{1.3mm} = 2.25 $ g/cm$^{-2}$ and 
is $\kappa_{1.3mm} = 0.4 $ g/cm$^{-2}$. 
The solid mass of the streamer thus amounts 
%to M$_d =$ 43 $\pm$ 5 M$_{\oplus}$ when considering a$_{max} \gtrsim$ 100 $\mu$m, or rather 
to M$_{dust}^{streamer} = $ 245 $\pm$ 25 M$_{\oplus}$.
% if a$_{max} \leq$ 1 $\mu$m. 
On the other hand, considering usual assumptions for protoplanetary disks to derive the mass of the inner region (e.g. \citealt{Andrews2013}, \citealt{Pascucci2016}, \citealt{Ansdell2016}), i.e. a temperature of 20 K and an absorption opacity $\kappa_{1.3mm} =$ 2.25 g/cm$^{-2}$, we obtain M$_{dust}^{inner} = $
58 $\pm$ 6 M$_{\oplus}$, given the integrated flux in Tab. \ref{tab:gaussfit}. %in the high opacity case or 313 $\pm$ 32 M$_{\oplus}$ in the low opacity one.
The reported uncertainties on the derived masses only reflect the dominant flux calibration error of 10\%. We note that since the compact, inner emission is optically thick, mass estimations for the inner region can be wrong by a factor up to $\sim$5 (\citealt{Ballering2019}, \citealt{Ribas2020}, \citealt{Macias2021}, \citealt{Xin2023}, \citealt{Rilinger2023}) and that further underestimation of the millimeter-derived mass can be due to scattering effects in the very inner, optically thick region (\citealt{Zhu2019}, \citealt{Liu2019}). 
Taking all of this into account, we find that the mass of the streamer seems to exceed or be comparable to the mass of the inner region. Given the resolution of the observations, this mass does not correspond to the mass of the protoplanetary disk alone but also accounts for contribution from the inner infalling material. 

Finally, we can constrain the infall rate onto M512 and its disk. Considering the typical free-fall timescale, t$_{ff} \propto \sqrt{{r^3}/{G M}}$, a particle with null initial velocity would fall from 4000 au onto the 0.15 $M_{\odot}$ protostar in roughly 50 kyr. This estimate is consistent with the TIPSY outputs ran in Sect. \ref{sec:kinematics}.
Thus, assuming the whole streamer will infall onto the inner system, and a typical dust-to-gas ratio of 0.01, the total mass infall rate 
%in the two extreme cases we considered would be $\dot M$ =  2.6 $\cdot$ 10$^{-7}$ M$_{\odot}$/yr or 
is $\dot M$ =  1.5 $\cdot$ 10$^{-6}$ M$_{\odot}$/yr. 

However, it is imperative to bear in mind that the infall rate may be lower if the streamer will not interact in its entirety with the inner protoplanetary disk. Consequently, while we undertake a discussion of the implications of our study under the assumption that the entire streamer will ultimately infall on the inner disk, we emphasize the sensitivity of these outcomes on the mass that will actually be captured by the source. Observations with higher spatial and spectral resolution are needed to clear the remaining sources of uncertainty in the analysis: the geometry of the inner system, the line-of-sight velocity of the star, and the impact of the infall via shock tracers such as SO (e.g., \citealt{Garufi2022}).
%new paragraph about comparison 
It is noteworthy that outbursts observed for young stellar objects have been proposed to be triggered by massive accretion events \citep[e.g.,][]{Bonnell1992, Aspin2003}. M512 might represent a rare case for which we witness a late infall event that is delivering mass at a significantly high rate.  
%In fact, considering the free-fall timescales and mass infall rates of which above, a total of 0.075 M$_{\odot}$ might build up in the disk. 
Assuming infall of the entire structure, and outbursts durations of $\sim$100 yr, this rates would imply a possible future accretion event of 7.5$\cdot$10$^{-4}$ M$_{\odot}$/yr, or a series of smaller ones. For example, if accretion would build up material in the disk for 10 kyr\footnote{The typical duration interval between subsequent outbursts for FUOr events \citep{Scholz2013}.}, this would imply FUOr-like accretion ($\sim$10$^{-4}$ M$_{\odot}$/yr, \citealt{Kenyon1988}), which might then cause outbursts. The accretion rate of M512 has been characterized to be in a quite ordinary accretion state, i.e. L$_{acc} <$ L$_*$, with $\dot M \sim 2 \cdot 10^{-7}$ M$_{\odot}$/yr, based on single epoch observations \citep{Caratti2012}. However, M512 has been classified as a bursting source based on multi-epoch WISE 3.4$\mu$m and 4.6$\mu$m photometry by \citet{Park2021}. Over the course of 6.5 yr and with a time resolution of 0.5 yr, they observed three sharp increases of the flux by a factor $\sim$2.5 in both bands.  This source represents an exciting testbed to explore the possible link between infalling streamers and accretion outbursts.

Furthermore, the mass infall rate onto M512 is similar to the one used by, e.g., \citet{Bae2015} and \citet{Kuznetsova2022} in their simulations, in which they found that material infalling at these high rates on protoplanetary disks has the power to drive significant substructures. M512 thus represents a unique source to probe whether infall can drive substructures, which are often regarded as birthplaces of planets.

We note that an additional source of uncertainty for the derived masses comes from the choice of dust opacities. While we have so far considered DSHARP opacities, other works have made use of different prescriptions, such as the DIANA opacities \citep{Min2016} or opacities constrained from solar system observations \citep{Pollack1994}. These alternative choices would imply absorption opacities systematically larger in the submillimeter regime, up to $\sim$3 or 10 times the DSHARP ones, respectively. Thus the mass derived through them would be up to ten times lower both for the streamer and the inner disk. We point out that consistent choices of opacities imply the same ratio between the streamer and the disk's mass.

\subsection{Origins and frequency}
\label{sec:origins}
The origin of streamers is still a matter of debate. Some authors have proposed that they are a channelled inflow of material from protostellar envelopes or even from beyond the prestellar core scales (e.g., \citealt{Yen2014}, \citealt{Yen2019}, \citealt{Pineda2020}). The extended emission in the surroundings of M512 discussed here, however, is not reminiscent of irregular envelopes and it is infalling in a seemingly ordered, arc-like manner (see Sect. \ref{sec:kinematics}). Moreover, CO observations seem to indicate that M512 lost most of its initial envelope \citep{grant2021}. Other explanations involve the interaction of M512 with closeby objects, such as a flyby with another stellar object or interactions with close interstellar clouds. 

%A close encounter with another stellar object would be consistent with the observed (Appendix \ref{app:moment}) velocity gradient around M512 (\citealt{Clarke1993},  \citealt{Cuello2019}, \citealt{Vorobyov2020}). 
Flyby events might be quite common in star-forming regions, especially during the protostellar stages, where stars form close to each other in clusters (\citealt{Pfalzner2013}, \citealt{Winter2018b}, \citealt{Lebreuilly2021}, \citealt{offner2022}). In this sense, the morphology of M512 is reminiscent of the ones observed for RW Aurigae (\citealt{Cabrit2006}, \citealt{Dai2015}, \citealt{Rodriguez2018}), AS 205 \citep{Kurtovic2018}, UX Tau \citep{menard2020}, and Z CMa \citep{Dong2022}), binary star systems for which a long molecular arm extends from the primary star and points toward a secondary object. 
%We checked for the presence of nearby sources for a recent flyby event. \citet{grant2021} reported a source 50$\farcs$0 (21,000 au) north-west of our target. However, no Gaia measurements are available for this source, thus impeding discussions about a possible encounter at present. We report on the presence of a second source (2MASS 05401156-0730409) in Appendix \ref{app:flyby}. This is at a distance of 100$\farcs$0 (41,000 au), and Gaia DR3 distance and proper motions are available. 
In Appendix \ref{app:flyby} we discuss how a close encounter might have occurred between M512 and another source a few hundreds kyr ago. This is much longer than most flyby events for which we still observe stripped material. Indeed, simulations suggest that the re-circularization of material happens in a few kyr (e.g., \citealt{Cuello2023}), making the observation of such an event extremely unlikely. Moreover, spirals excited due to stellar flybys are usually of the same order of magnitue of the outer radius of the stripped disk ($\lesssim$ 3 R$_{out}$, \citealt{Smallwood2023}), while the arc we observe extends for thousands of astronomical units. The flyby scenario is thus hard to reconcile with our observations. We expand on our arguments in Appendix \ref{app:flyby}. 

The velocity gradient we observe in the arc of M512 (see Fig.\ref{fig:moment1}) is also consistent with the capture of material from a nearby cloud with a non-zero initial velocity. The material is then accelerated toward the central star. \citet{Scicluna2014} and \citet{Dullemond2019} explored the possibility that forming protostars might capture material from nearby small clouds as they travel within their natal environment. In this scenario, due to its non-null initial angular momentum, the material does not fall directly on the protostar but feeds the disk, or even forms a new one. This capture process has been proposed to explain the observed infrared excess around very evolved, $\geq$10 Myr old stars \citep[e.g.,][]{Beccari2010, DeMarchi2013a, DeMarchi2013b} and the $M_* - \dot M$ correlation in pre-main-sequence stars \citep[e.g.,][]{Padoan2005, ThroopBally2008, Padoan2014}. Examples of objects that might be undergoing such scenarios are AB Aur \citep{Nakajima1995, Grady1999} and HD~100546 \citep{Ardila2007}. The hydrodynamics simulations of \citet{Dullemond2019} and \citet{Kuffmeier2020} demonstrate how the capture of cloud fragments would lead to the formation of arc-shaped structures, much like the one observed in M512 (\citealt{Scicluna2014}, \citealt{Dullemond2019}). Additionally, \citet{Scicluna2014} computed the probability of cloudlet capture by a stellar object in different environment density conditions. They found that for a region in which dense clumps occupy a fraction f$_V$ of the total volume, the number of stars that one expects to observe as cloudlet capture accretors at a given time is an order of magnitude larger than the volume filling-factor of dense clumps. Thus, if dense clumps only occupy a volume fraction as small as f$_V =$10$^{-4}$, we expect one in a thousand objects to show late accretion. We note that M512 is one of three objects showing extended continuum emission in the Survey of Orion Disks with ALMA (SODA) \citep{vanterwisga2022}, that includes 873 objects. The overall median accreted mass in all the simulation grids they explored is 0.01$M{_\odot}$. They also demonstrated that the structures formed due to capture can last even for 10$^4$-10$^5$ yr (see their Section 2). These values are comparable to M512's streamer mass and free-fall timescale.
Furthermore, we note that dust grains in ISM clouds should not exceed a few tenths of microns (e.g., \citep{Mathis1977}), thus further supporting a scenario for which the $\beta \sim 1.6$ measured in Sect. \ref{sec:discuss_dust} indicates ISM-like grains. Larger-scale maps than available at present would be helpful in linking the streaming material around M512 with the neighboring environment, and to understand whether the observed streamer presented in this work is replenished by the surrounding environment or represent the full extent of the mass that will be delivered on the disk of M512.  

Lastly, continuum emission has sometimes been detected along the cavity walls of protostellar outflows where, due to higher temperatures, the dust might be brighter (e.g., \citealt{Maury2018}, \citealt{LeGouellec2023}). In the case of M512, however, the arc-like emission is quite bent and not reminiscent of more cone-like outflow cavities (e.g., \citealt{garufi2021}, \citealt{hsieh2023}). Additionally, CO emission is optically thick and the C$^{18}$O line profile indicates low velocities, about 500-700 m/s, and we observe both redshifted and blueshifted emission within the same arc (Fig. \ref{fig:moment1}). Finally, M512 is a class I, very low mass star, implying such a large outflow would be unlikely. It is thus not straightforward to reconcile our observations with the outflow scenario. We also note that our mass estimates would still be valid with the caveat that, at the cavity wall of a low velocity outflow, temperatures would be slightly higher than what we assumed for the cold streamer case (e.g., \citealt{Droz2015}, \citealt{Flores-Rivera2021}). This would imply a mass a factor of a few lower, and thus still a significantly high dust mass (M$_{dust}$ $\gtrsim$ 50 M$_{\oplus}$). Considering typical class I outflows mass loss rates of a few 10$^{-8}$ M$_{\odot}$/yr \citep{Fiorellino2021}, it would take over 1 Myr yr to lift M$_{dust}$ $\gtrsim$ 50 M$_{\oplus}$, while the mass loss rates quickly decline over the first $\sim 10^5$ yr of star formation. We thus tentatively discard this scenario. However, non-isotropic accretion from the envelope to the central source has been suggested to happen along the cavity walls of outflows (\citealt{LeGouellec2019}, \citealt{Cabedo2021}).
Observations of bright and optically thin outflow tracers will help clear out this case together with higher resolution observations of the central source to resolve the geometry of the disc and thus the direction along which an outflow could be launched.

In summary, given the constraints of Sect. \ref{sec:kinematics}, and the qualitative arguments laid in this section, we reinforce our suggestion that the arc-like structure around M512 is an infalling streamer, possibly caused by a cloudlet capture event.

\subsection{Implications for planet formation}
\label{discuss:implications}

Streamers can have significant implications for planet formation and protoplanetary disk evolution. M512 represents a unique case for which dust continuum emission is detected in two ALMA bands, thus enabling us to better constrain its dust properties and mass content. The structure we observe is delivering a substantial amount of mass to the inner disk of M512, carrying several significant implications that we discuss in what follows. In what follows, we assume the whole streamer will infall on the central disk, consistently with the bound solutions found in the kinematical analysis we lay out in Sect. \ref{sec:kinematics}. However, it is possible that only a fraction of the detected structure will infall on the inner system, thus the infall rate could be a factor of a few lower.

To begin with, the streamer is replenishing the system with an amount of mass that exceeds or is comparable to the mass of the protoplanetary disk. If the infalling dust were captured by a nearby cloudlet where maximum grain sizes are 1-10 $\mu$m (see \ref{sec:discuss_dust}), a total mass of up to 0.075 M$_{\odot}$ may infall on the disk of M512. The infall can potentially double the available mass for planet formation in the system. The mass budget problem, i.e. the apparent lack of necessary mass to form known exoplanetary systems starting from the class II protoplanetary disks we observe in the Galaxy (e.g., \citealt{Testi2016}, \citealt{Manara2018}, \citealt{Williams2019}, \citealt{Sanchis2020}, \citealt{Tychoniec2020}), could thus be partially bridged by late infall of material on evolved disks. Additionally, the supply of material around low mass stars such as M512 could represent a possibility for the formation of gas giants around these kind of objects. While such exoplanets are, at least rarely, detected (e.g., \citealt{Morales2019}, \citealt{Bryant2023}), their formation pathways remain unclear \citep{Liu2020}.

Additionally, M512 represents a candidate to study the origin of substructures in evolved disks. In fact, infall of material onto disks has been proposed to be among the mechanisms that can trigger such substructures \citep{Thies2011, Hennebelle2017, Kuffmeier2018, Ginski2021}. \citet{Bae2015} and \citet{Kuznetsova2022} demonstrated how infall can drive the formation of rings and gaps in disks with infall rates consistent with what we find for M512. 
A possible instance of this process 
was detected by \citet{Segura-Cox2020}, who observed rings and gaps in in the protoplanetary disk around Oph IRS63, recently shown to be subject to anisotropic infall \citep{Flores2023}. Another example is HL Tau, where both disk structures and an infalling streamer were observed (\citealt{AlmaPartenership2015}, \citealt{Yen2019}). Studying the impact of streamers on disks that display substructures can inform us about the necessity, or lack thereof, to invoke planets to carve such structures at early stages of star and planet formation. Moreover, the formation of substructures and the induction of turbulence in the disk can alter the timescales of grain growth since clumping of grains is more efficient where these conditions are present. Higher-resolution observations of M512 will be key to unveil the geometry of its inner disk and search for potential substructures.

Lastly, the infalling material will impact the disk at some radius, and will potentially shock with its surroundings. In the vicinity of the shock, the disk temperature will rise \citep{Garufi2022}. Thus, the streamer will change the physical and chemical properties of the disk, as shown for e.g, HL Tau in \citep{Garufi2022}. M512 is known to undergo outbursts \citep{Park2021} that can greatly alter the physical and chemical properties of protoplanetary disks influencing both the evolution of dust \citep{Houge2023} and gas \citep{Owen2014, Wiebe2019}. Whether these are linked to the infalling streamer or not will be an exciting matter for further studies.

\section{Conclusions}
\label{sec:conc}
We have presented new ALMA B3 observations of M512 in the Lynds 1641 region of the Orion A molecular cloud and combined them with archival ALMA B6 data to constrain the dust properties of a 4,000 au arc-like structure extending from the central source. We find that:

\begin{itemize}
    \item[$\bullet$] The structure is characterized by a velocity gradient consistent with infall (see Fig.~\ref{fig:moment1}). The morphology of the extended emission and the velocity gradient are consistent with the capture and infall of material from a nearby cloudlet. 

    \item[$\bullet$] The compact ($\sim$ 600 au) region surrounding M512 is optically thick at the wavelengths considered here ($\alpha\sim2$). Higher resolution observations will be key to study this inner region and constrain its geometry and the impact of the infall event, which we cannot probe with current observations.
    
    \item[$\bullet$] The streamer is optically thin with $\alpha_{S}\sim3.2$ along the whole structure. Accounting for the low temperatures expected at its scales, we obtain a dust opacity spectral index $\beta$ = 1.62$\pm$0.04, typical of ISM-like dust (Sect. \ref{sec:discuss_dust}).
    
    \item[$\bullet$] We constrain the dust mass of the streamer to be $\sim 250 M_{\oplus}$\footnote{This value is obtained when considering the opacities for ISM-like grains of \citet{Birnstiel2018}, thus $\kappa_{1.3mm} = 0.4 $ g/cm$^{-2}$, and the temperature profile give in Eq.\ref{eq:temp}.} (see Sect. \ref{discuss:mass}). Given typical free-fall timescales, the streamer could be delivering up to 1.5 $\cdot$ 10$^{-6}$ M$_{\odot}$/yr onto the inner disk and last about 50 kyr. We note this estimates assume the whole structure will infall. 

    \item[$\bullet$] We discuss the possible origins and impact that this event can have on the planet-forming disk of M512, including the replenishment of mass, the formation of substructures, and changes in the physical and chemical conditions due to shocks and/or accretion outbursts (Sect. \ref{discuss:implications}).
    
\end{itemize}

Streamers observations represent a new window on star and planet formation, reminding us how this process is not isolated and is highly dynamic within star forming regions. Further observations will be necessary to both link streamers to their potential large-scale progenitors and test the impact they have on planet-forming disks on the smaller scales. M512 is a quite unique source that will play a crucial role in these studies. This target is a rare opportunity to study both gas and dust, their physical properties and dynamics, thus enabling an all-around comparison to state-of-the-art models.

\begin{acknowledgements}
This work was partly supported by the Italian Ministero dell’Istruzione, Universit\`{a} e Ricerca through the grant Progetti Premiali 2012-iALMA (CUP C52I13000140001), by the Deutsche Forschungsgemeinschaft (DFG, German Research Foundation) - Ref no. 325594231 FOR 2634/2 TE 1024/2-1, by the DFG Cluster of Excellence Origins (www.origins-cluster.de). This project has received funding from the European Union’s Horizon 2020 research and innovation program under the Marie Sklodowska- Curie grant agreement No 823823 (DUSTBUSTERS) and from the European Research Council (ERC) via the ERC Synergy Grant ECOGAL (grant 855130). Funded by the European Union (ERC, WANDA, 101039452). Views and opinions expressed are however those of the author(s) only and do not necessarily reflect those of the European Union or the European Research Council Executive Agency. Neither the European Union nor the granting authority can be held responsible for them. This work benefited from the Core2disk-III residential program of Institut Pascal at Université Paris-Saclay, with the support of the program “Investissements d’avenir” ANR-11-IDEX-0003-01. This paper makes use of the following ALMA data: ADS/JAO.ALMA\#2019.1.01813.S, ADS/JAO.ALMA\#2019.1.00951.S, ADS/JAO.ALMA\#2022.1.00126.S. ALMA is a partnership of ESO (representing its member states), NSF (USA) and NINS (Japan), together with NRC (Canada), MOST and ASIAA (Taiwan), and KASI (Republic of Korea), in cooperation with the Republic of Chile. The Joint ALMA Observatory is operated by ESO, AUI/NRAO and NAOJ. We thank the entire ALMA team for their dedication to provide us with the data we used for this work. We thank Nicolas Cuello for helpful insights and discussion. We thank the referee for their insightful comments, which helped us to improve the quality of this work.
\end{acknowledgements}


\bibliographystyle{aa} % style aa.bst
\bibliography{biblio} % your references Yourfile.bib

\appendix

\section{Datasets}
We report a summary of the datasets utilized in this work in Table \ref{table:1}. 

\begin{table*}[t]
\footnotesize
\caption{Summary of the datasets that have been used in this work.}             
\label{table:1}      
\centering          
\begin{tabular}{c c c c c c c}     % n columns 
\hline      
                      % To combine 4 columns into a single one 
Project Code & P.I. & Date &  Integration (s) & Resolution (arcsec) & Frequency (GHz) &  CASA version  \\ 
\hline 
  \hline \\
    & & & Band 3 & & &   \\    
     2022.1.00126.S & Wendeborn, J. & 2023/01/16 &  7529  & 1.05 & 89-105 & 6.4.1-12 \\ 
     
\hline \\
& & & Band 6 & & &  \\
    2019.1.01813.S & van Terwisga, S. & 2019/11/27 & 32 & 1.13  & 223-243 & 5.6.1-8 \\  
    2019.1.00951.S & Grant, S, & 2019/12/12 & 72  & 1.15 & 216-233  & 5.6.1-8 \\ 
\hline      
                                    
\end{tabular}
\end{table*}

% \section{Escape velocity maps}
% \label{app:escape}
% We produced maps of the ratio between the radial velocity of the detected C$^{18}$O with the escape velocity. To asses whether the material in the streamer is bound to the inner protostar, we considered M = 0.15 $M_{\odot}$, v$_{sys}$ = 3.6 km/s and different inclinations. We reported the case for the most probable inclination in the sky ($\sim$60$^{\circ}$) in Sect. \ref{sec:kinematics}. Here, we report the cases where \textit{i}=30$^{\circ}$, 45$^{\circ}$ (Figures \ref{fig:30deg}, \ref{fig:45deg}).
% \begin{figure}[t]
%     \centering
%     \includegraphics[width=\linewidth]{escape_map_inc30.png}
%     \caption{Same as Fig. \ref{fig:escape60}, but with an assumed inclination of 30 degrees.}
%     \label{fig:30deg}
% \end{figure}
% \begin{figure}[t]
%     \centering
%     \includegraphics[width=\linewidth]{escape_map_inc45.png}
%     \caption{Same as Fig. \ref{fig:escape60}, but with an assumed inclination of 45 degrees.}
%     \label{fig:45deg}
% \end{figure}

% The moment-one map shows a velocity gradient from the outer to the inner streamer. This morphology is consistent with infall. A thorough kinematics analysis of this object will be presented in Gupta et al. (in prep.).

% \begin{figure}
%     \centering
%     \includegraphics[width=\linewidth]{M512_moment_map_C18O.png}
%     \caption{Moment one map of $^{18}$CO J $= 2-1$ transition around M512 ($>2 \sigma$). The black contours are the B6 continuum [3, 10, 100]$\sigma$ levels.}
%     \label{fig:moment1}
% \end{figure}

\section{Compact emission modelling}
\label{app:gaussfit}
In order to measure $\alpha$ and $\beta$ along the streamer free of flux contamination from the inner optically thick region, we have subtracted the compact contribution centered on the source, in two ways. 

\begin{figure*}
    \centering
    \includegraphics[width=0.65\linewidth]{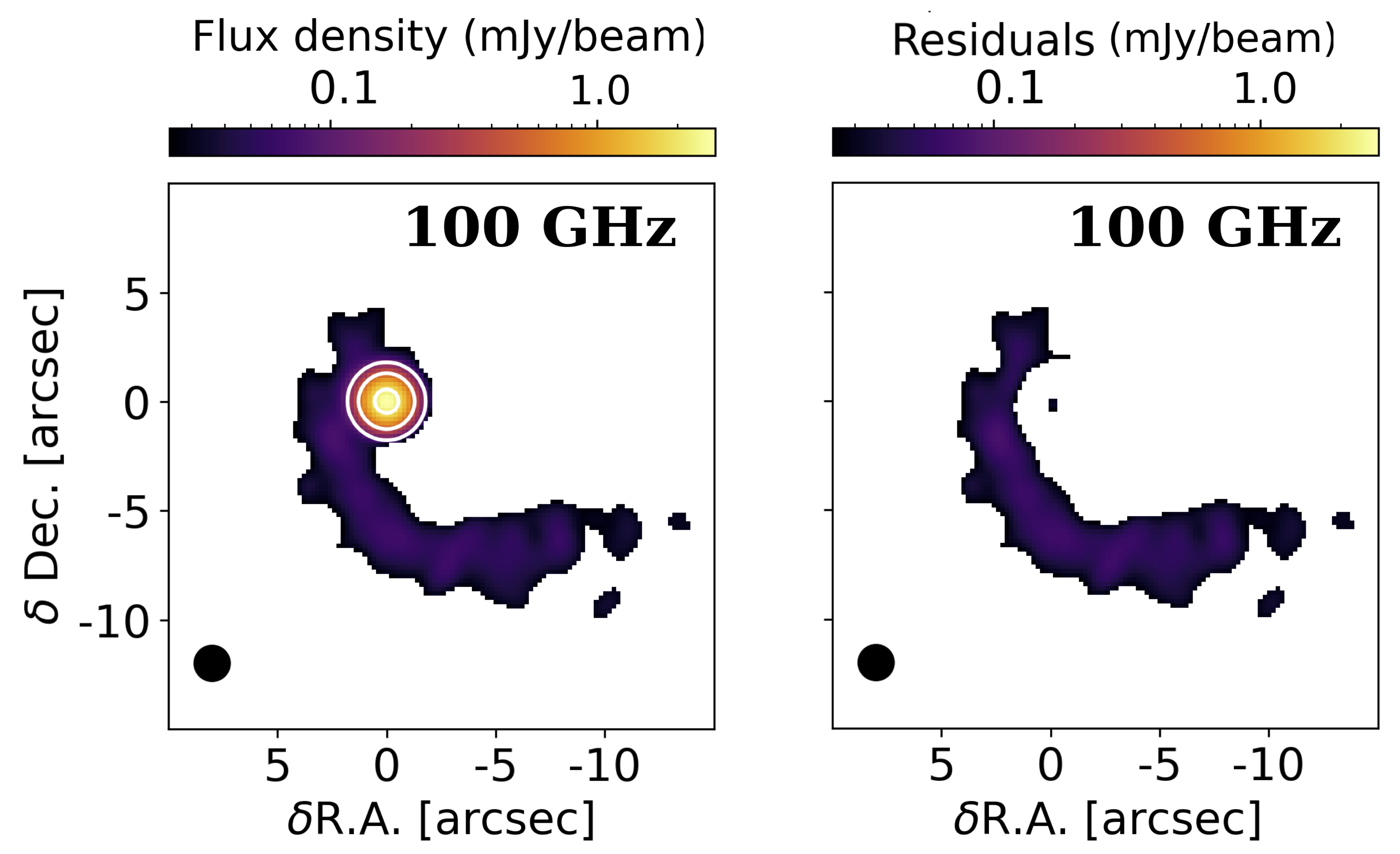}
    \caption{Contour levels (white) of the Gaussian model on top of the ALMA B3 (100 GHz) data of M512 (left). The colorbar reports the flux values of the original image. The residuals of the model are shown on the right and their colorbar lays in the same range of values of the original emission to help the comparison.}
    \label{fig:gauss_fit3}
\end{figure*}

\begin{figure*}
    \centering
    \includegraphics[width=0.65\linewidth]{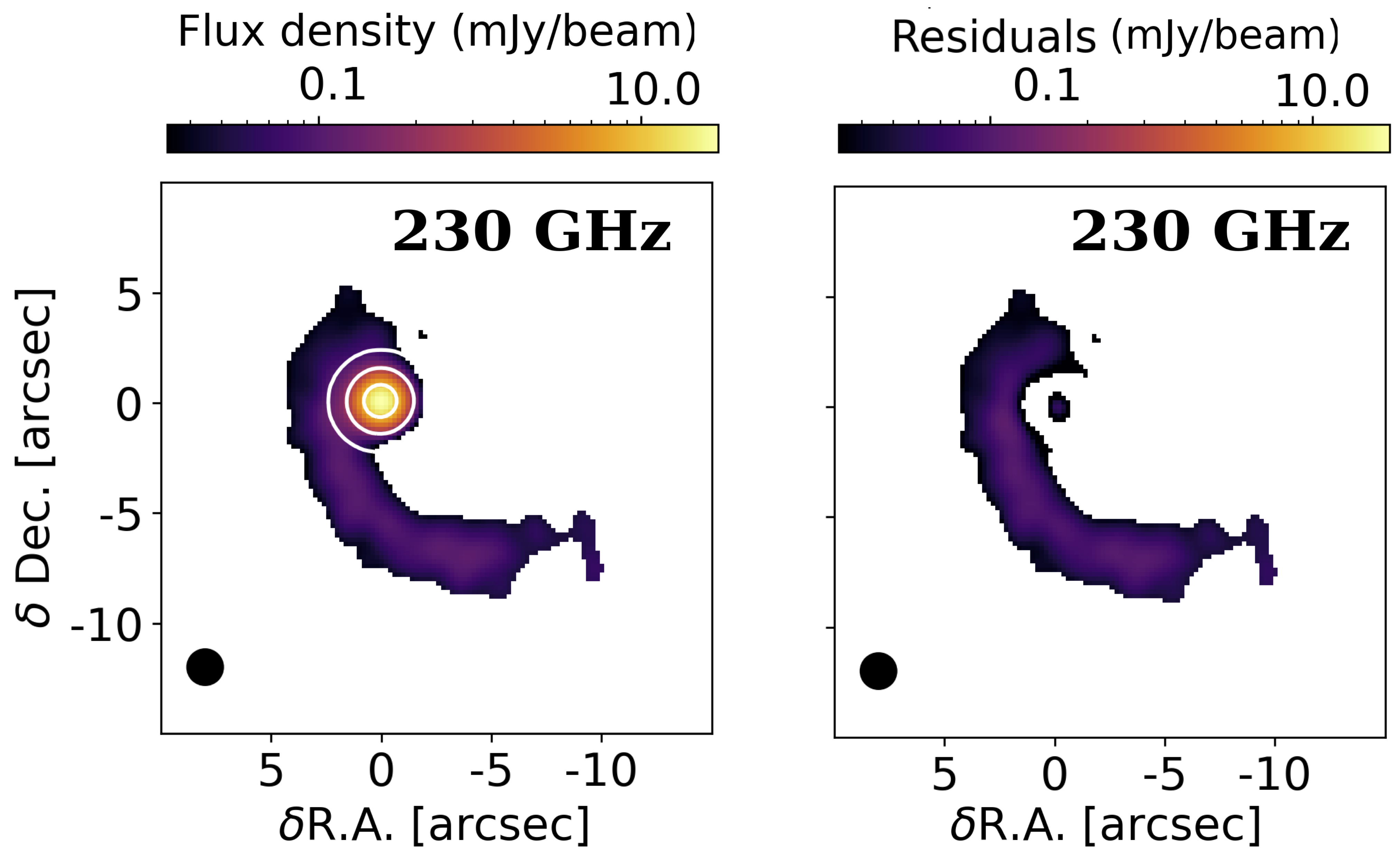}
    \caption{Same as Fig.~\ref{fig:gauss_fit3} but for B6 (230 GHz) data.}
    \label{fig:gauss_fit6}
\end{figure*}

\begin{figure}
    \centering
    \includegraphics[width=\linewidth]{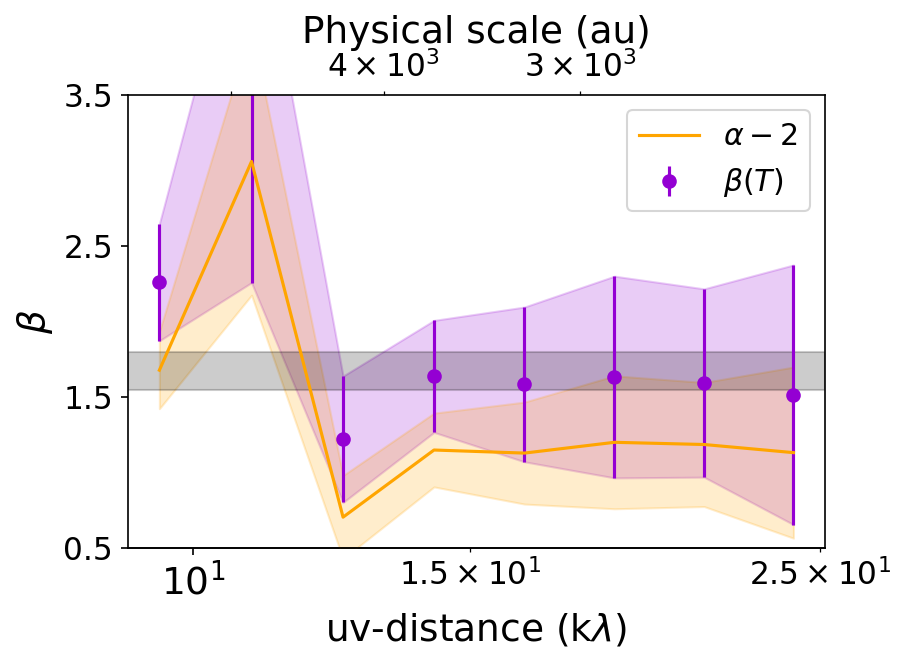}
    \caption{Spectral (orange) and dust opacity (violet) indices as measured across probed uv-distances, after subtraction of a point-like source. The mean values $\alpha \sim 3.1$ and $\beta \sim 1.6$ are consistent with what found across the scale of the streamer in the image plane analysis.}
    \label{fig:betauv}
\end{figure}

First, we fitted a gaussian model to the central region in the image plane. Figures \ref{fig:gauss_fit3} and \ref{fig:gauss_fit6} show the data, the model, and the residuals of the fit. Table \ref{tab:gaussfit} summarizes the fitted parameters for the models in the two bands. The spectral index $\alpha$ and dust opacity spectral index $\beta$ have been measured once the optically thick region has been modelled out (Sect. \ref{sec:spectral_index}). 
As a robustness check, we have tested whether our results are robust against a different modelling scheme. We have used the \textit{uvmodelfit} CASA routine to fit a point source component to the visibilities, since the disc is unresolved at the resolution of our observations. The function fits for an offset from the phase center and the flux of the point source. We find the same offset from center as for the gaussian fit and a total integrated flux density of 16.1 mJy in B6 and 2.6 mJy in B3.
We then subtracted this contribution from the visibility amplitudes (A = $\sqrt{Re^2 + Im^2}$) at all scales, and binned them. In each bin, the uncertainty is given by the combined uncertainties on the real and imaginary parts of the visibilities, together with the ALMA calibration error (1$\sigma$ = 5\%, \citealt{Remijan2019}).
We finally computed the spectral and dust opacity indices using Eq. \ref{eq:alpha} and \ref{eq:beta} as a function of uv-distance after the subtraction, as shown in Fig. \ref{fig:betauv}. A bump at the short uv-distances is caused by the combination of the asymmetric shape of the streamer with the uv-coverage of the observations. The large scatter in amplitude at these scales also results in a larger error around the mean and thus on a larger uncertainty on $\alpha$ and $\beta$. The very first bin (at the shortest common uv-distance) shows $\beta \sim 2$, because of a raise in the B6 visibility amplitudes that is not observed in B3 due to uv-coverage. Shorter baselines, and thus larger recoverable scales observations, are needed to more robustly measure $\beta$ in the outskirts of the streamer. The higher $\beta$ would still be in agreement with the conclusion of small, ISM-like grains.
Finally, both azimuthally averaged spectral indices are consistent with what found in the image plane: $\alpha_S = 3.1 \pm 0.1$, and $\beta = 1.55 \pm 0.05$.

\begin{table*}[h]
\footnotesize
\caption{Results of the Gaussian fit to the compact unresolved region around M512.}             
\label{tab:gaussfit}      
\centering          
\begin{tabular}{c c c c}    \hline
Band & Shift from source coordinates (arcsec) & Integrated flux density (mJy) & $\sigma_x$ x $\sigma_y$ (arcsec) \\ \hline \hline \\    
 3   &    (0.060 $\pm$ 0.003, 0.080 $\pm$ 0.003)   &  2.8 $\pm$ 0.1   & (0.696 $\pm$ 0.002) x (0.691 $\pm$ 0.002) \\
 6   &     (0.020 $\pm$ 0.008, 0.020 $\pm$ 0.008)   &  19.4 $\pm$ 0.2    & (0.740 $\pm$ 0.008) x (0.723 $\pm$ 0.007) \\ \hline 
\end{tabular}
\end{table*}

\section{Any flyby candidates?}
\label{app:flyby}
\begin{figure}
    \centering
    \includegraphics[width=\linewidth]{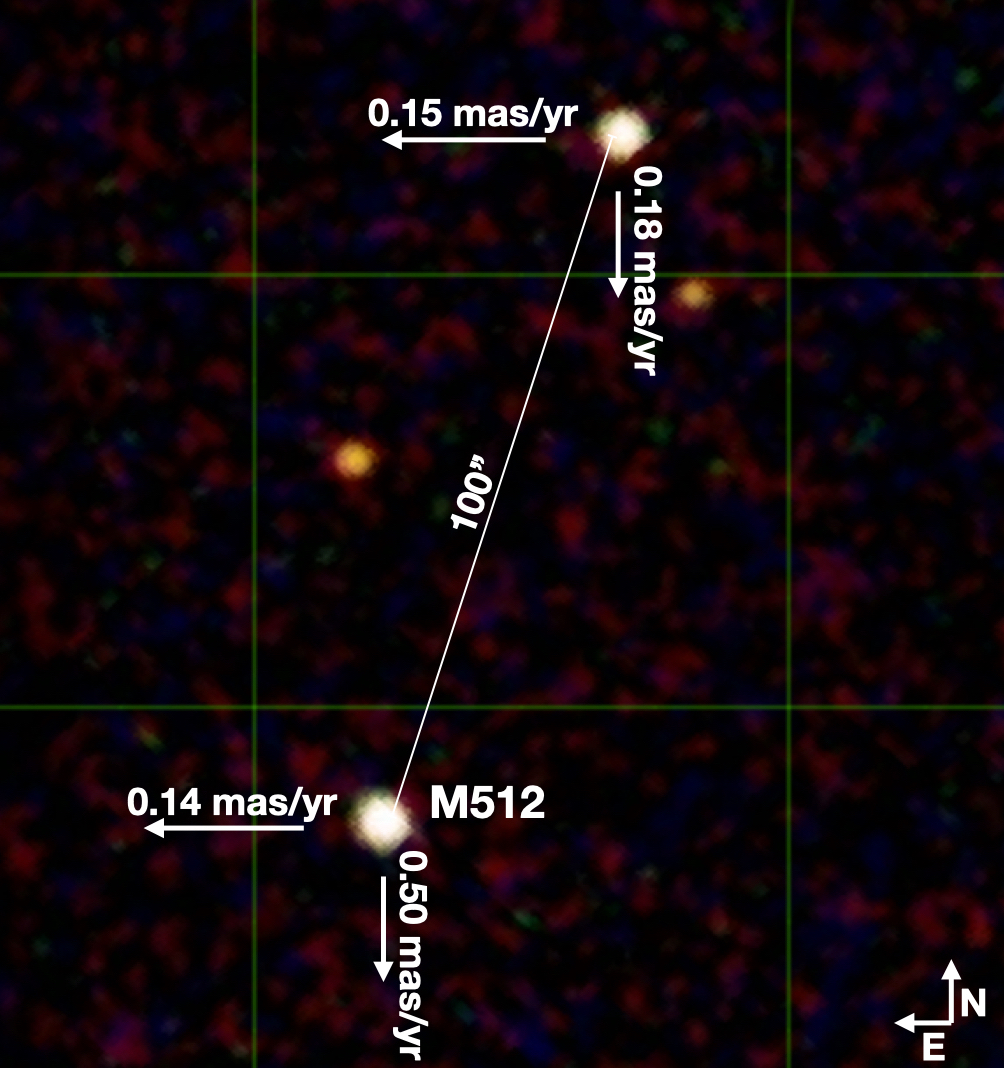}
    \caption{2MASS map of the surroundings of M512, whose arc extends westwards. 
    A young stellar object (2MASS J05401156-0730409) is at a projected distance of approximately 41,000 au from M512. The Gaia DR3 mean proper motion components are overlaid as arrows and their magnitude is reported in milliarcseconds per year.
    A close encounter might have happened, but only a few hundreds kyr ago. Each box of the grid is 1'x1'.}
    \label{fig:flyby}
\end{figure}

We here note the presence of a class II YSO ($\alpha_{2-25\mu m} = -1$; \citealt{Megeath2012}) at projected distance of 100$\farcs$ (41,000 au) from M512. This source, 2MASS J05401156-0730409, was studied in continuum by \citet{vanterwisga2022}, who reported a integrated 1.3mm flux of 1.65 $\pm$ 0.06 mJy and a dust mass estimate of 8.1 $\pm$ 0.3 M$_{\oplus}$, assuming optically thin emission, a dust temperature of 20 K, and a dust absorption opacity of 2.3 cm$^2$/g.
Gaia DR3 \cite{Gaia2022} reported a parallax of 2.54 $\pm$ 0.18 for the latter, indicating it is at a distance consistent (however different) with M512, whose Gaia DR3 parallax is 2.37 $\pm$ 0.12. The secondary object has a measured Gaia DR3 proper motion in declination of pmDec $= -$0.18 $\pm$ 0.15 mas/yr and a proper motion in right ascension of pmRA $=$ 0.14 $\pm$ 0.16 mas/yr. M512 moves in the sky with pmDec $= -$0.50 $\pm$ 0.10 mas/yr and pmRA $=$ 0.15 $\pm$ 0.11. 
Considering that the sources have $-$0.32 $\pm$ 0.18 mas/yr relative motion in declination and $\delta$pmRA = 0.01 $\pm$ 0.19 in right ascension.
%, M512 could have had a close encounter with this other source between 200 and 700 kyr ago. 
The relative proper motions are consistent with a close encounter the two objects between 200 and 700 kyr ago (1$\sigma$ range), on the plane of the sky. However, considering the error bars on the Gaia proper motion and parallax measurements, it's possible that no flyby occurred at all. We report the 2MASS sky map in Figure \ref{fig:flyby}. 

Both spatial and time scales yield major counter arguments to this scenario. For a disc of size R$_{disc}$ perturbed by a star during a close encounter R$_{flyby}$ $\sim$ R$_{disc}$, simulations suggest that induced arc-like spirals would have a spatial extent of the order of three times R$_{disc}$ at most (see \citet{Cuello2019}). The streamer around M512, however, is 4,000 au or longer (depending on projection effects and interferometric filtering), and would require an unreasonably extended protoplanetary disk that we would have resolved with our observations.
Additionally, even in the earliest case compatible with Gaia constraints, a close encounter could have occurred between 200-700kyr ago, within 1sigma confidence. The typical free fall timescale t$_{ff} \propto \sqrt{r^3/Gm}$ for a particle infalling from 4,000 au onto a 0.15 M$_{\odot}$ star is roughly 50 kyr, thus the material should have already fallen onto the star. If the observed streamer were a flyby-induced spiral, its survival time would be at most a few times the orbital period at the outer edge of the disc \citep{Smallwood2023}, thus it would have been dispersed in a few kyr, making its observation very improbable. 
Lastly, the derived masses for the streamer around M512 and its inner region suggests that a flyby should have stripped away a significant portion of the disk (see Sect. \ref{discuss:mass}).
\citet{Pfalzner2005} demonstrated how even encounters with a similar mass object and with a periastron of the order of the disk radius would only strip up to about 50\% of the perturbed disk's mass\footnote{Although this argument relies on the systematically uncertin dust mass constraints.}. Finally, we inspected ALMA archival data for this nearby source (ID 2019.1.01813.S) and found no evidence for extended emission. 

As a cautionary argument, however, we note that flybys for which the stripped material has been observed in a structure exceeding 3 times R$_{disc}$ (RW Aur), and for which re-circularization timescales seem to exceed a few kyrs (HV Tau and DO Tau) have been observed in \citet{Rodriguez2018} and \citet{Winter2018}, respectively.
Overall, we suggest that a flyby of M512 with 2MASS J05401156-0730409 appears harder to reconcile with the observed extended and massive arc than the streamer scenario, but cannot be completely ruled out at this stage.

\end{document}